\documentclass[graybox, envcountchap]{svmult}

\usepackage{graphicx}        
\graphicspath{
	{book/}
	{book/chapter-introduction/}
	{book/chapter-bots/}
	{book/chapter-collateral/}
	{book/chapter-dataintenive/}
	{book/chapter-emotionanalysis/}
	{book/chapter-libraries/}
	{book/chapter-iac/}
	{book/chapter-mde/}
	{book/chapter-promisesandperils/}
	{book/chapter-socialforks/}
	{book/chapter-softwareheritage/}
	}                             

\usepackage[sectionbib]{chapterbib}

\usepackage{multicol}        
\usepackage[bottom]{footmisc}
\usepackage{enumitem}

\usepackage{newtxtext}       %

\usepackage{url}
\usepackage{pifont}

\usepackage{algorithm}
\usepackage{algpseudocode}
\usepackage{booktabs}
\usepackage{multirow}
\usepackage{wrapfig}
\usepackage{listings}
\usepackage{siunitx}
\usepackage{pgfplotstable}
\usepackage{array}
\usepackage{subcaption}
\usepackage{newfloat}
\usepackage{hyphenat}

\usepackage{pdflscape}
\usepackage{xspace}

\usepackage{diagbox} 

\newcommand{\chap}[1]{Chapter~\ref{#1:ch}}
\newcommand{\sect}[2]{Section~\ref{#1:sec:#2}}

\newcommand{\lst}[2]{Listing~\ref{#1:lst:#2}}

\usepackage[disable]{todonotes}

\newcommand*{\ie}{i.e.,\@\xspace}
\newcommand*{\eg}{e.g.,\@\xspace}
\newcommand*{\etal}{\emph{et~al.}\@\xspace}
\newcommand{\github}{{GitHub}\xspace}

\newcommand{\gitlab}{GitLab\xspace}

\newcommand{\gha}{GitHub Actions\xspace}
\newcommand{\actions}{GitHub Actions\xspace}

\newcommand\YAMLcolonstyle{\color{red}\mdseries}
\newcommand\YAMLkeystyle{\color{black}\bfseries}
\newcommand\YAMLvaluestyle{\color{blue}\mdseries}

\makeatletter

\newcommand\language@yaml{yaml}

\expandafter\expandafter\expandafter\lstdefinelanguage
\expandafter{\language@yaml}
{
  keywords={true,false,null,y,n},
  keywordstyle=\color{darkgray}\bfseries,
  basicstyle=\YAMLkeystyle,                                 
  sensitive=false,
  comment=[l]{\#},
  morecomment=[s]{/*}{*/},
  commentstyle=\color{purple}\ttfamily,
  stringstyle=\YAMLvaluestyle\ttfamily,
  moredelim=[l][\color{orange}]{\&},
  moredelim=[l][\color{magenta}]{*},
  moredelim=**[il][\YAMLcolonstyle{:}\YAMLvaluestyle]{:},   
  morestring=[b]',
  morestring=[b]",
  literate =    {---}{{\ProcessThreeDashes}}3
                {>}{{\textcolor{red}\textgreater}}1     
                {|}{{\textcolor{red}\textbar}}1 
                {\ -\ }{{\mdseries\ -\ }}3,
}

\lst@AddToHook{EveryLine}{\ifx\lst@language\language@yaml\YAMLkeystyle\fi}
\makeatother

\newcommand\ProcessThreeDashes{\llap{\color{cyan}\mdseries-{-}-}}

\usepackage{wasysym}

\DeclareFloatingEnvironment[
    fileext=los,
    listname={List of Listings},
    name=Listing,
    placement=!h,
    within=none,
]{listing}

\lstdefinestyle{docker}{
  language=bash,
  morekeywords={RUN,FROM,MAINTAINER},
  showstringspaces=false,
  frame=none,
}
\lstdefinelanguage{ansible}{
  morekeywords={name,vars,hosts,tasks,roles,role},
  keywordstyle=\bfseries,
  morecomment=[l][\textit]\#,
  morecomment=[s][\bfseries]{\{\{}{\}\}},
}
\lstdefinestyle{ansible}{
  language=ansible,
  basicstyle=\scriptsize\ttfamily,
}

\def\postbreak{%
  \raisebox{0ex}[0ex][0ex]{\ensuremath{\hookrightarrow\space}}}

\lstdefinestyle{searchstringstyle}{
	basicstyle=\ttfamily\footnotesize,
	breakatwhitespace=false,         
	breaklines=true,                 
	captionpos=t,                    
	keepspaces=true,                 
	numbers=none,                    
	numbersep=5pt,                  
	showspaces=false,                
	showstringspaces=false,
	showtabs=false,                  
	tabsize=2,
	frame=single
}

\definecolor{mauve}{rgb}{0.58,0,0.82}
\definecolor{dkgreen}{rgb}{0,0.6,0}
\definecolor{gray}{rgb}{0.5,0.5,0.5}

\definecolor{pblue}{rgb}{0.13,0.13,1}
\definecolor{pgreen}{rgb}{0,0.5,0}
\definecolor{pred}{rgb}{0.9,0,0}
\definecolor{pgrey}{rgb}{0.46,0.45,0.48}

\lstdefinelanguage{JavaScript}{
  keywords={break, case, catch, continue, debugger, default, delete, do, else, false, finally, for, function, if, in, instanceof, new, null, return, switch, this, throw, true, try, typeof, var, void, while, with},
  morecomment=[l]{//},
  morecomment=[s]{/*}{*/},
  morestring=[b]',
  morestring=[b]",
  ndkeywords={class, export, boolean, throw, implements, import, this},
  sensitive=true
}

\lstdefinestyle{JavaStyle}{%
  language=Java,
  showspaces=false,
  showtabs=false,
  breaklines=true,
  showstringspaces=false,
  breakatwhitespace=true,
  commentstyle=\color{pgreen},
  keywordstyle=\color{pblue},
  stringstyle=\color{pred},
  columns=flexible,
  basicstyle={\small\ttfamily},
  numbers=left,
  frame=tb,
  xleftmargin=2em,
  framexleftmargin=2em
}

\lstdefinestyle{JavaScriptStyle}{%
  language=JavaScript,
  showspaces=false,
  showtabs=false,
  breaklines=true,
  showstringspaces=false,
  breakatwhitespace=true,
  commentstyle=\color{pgreen},
  keywordstyle=\color{pblue},
  stringstyle=\color{pred},
  columns=flexible,
  basicstyle={\small\ttfamily},
  numbers=left,
  frame=tb,
  xleftmargin=2em,
  framexleftmargin=2em
}

\usepackage{tikz} 
\usetikzlibrary{shadows,calc}

\colorlet{innercolor}{black!60}
\colorlet{outercolor}{gray!05}

\pgfdeclarelayer{shadow} 
\pgfsetlayers{shadow,main}

\begin{document}

\title*{The \github Development Workflow Automation Ecosystems}
\author{Mairieli Wessel \and Tom Mens \and Alexandre Decan \and Pooya Rostami Mazrae}
\institute{Mairieli Wessel, Radboud University, Netherlands;
\email{mairieli.wessel@ru.nl}\\
Tom Mens, Alexandre Decan and Pooya Rostami Mazrae, University of Mons, Belgium.
Alexandre Decan is a Research Associate of the Fonds de la Recherche Scientifique - FNRS;
\email{firstname.lastname@umons.ac.be}
}
\maketitle
\label{WFA:ch}

\vspace{-2cm}
\abstract{
Large-scale software development has become a highly collaborative and geographically distributed endeavour, especially in open-source software development ecosystems and their associated developer communities. It has given rise to modern development processes (e.g., pull-based development) that involve a wide range of activities such as issue and bug handling, code reviewing, coding, testing, and deployment. These often very effort-intensive activities are supported by a wide variety of tools such as version control systems, bug and issue trackers, code reviewing systems, code quality analysis tools, test automation, dependency management, and vulnerability detection tools. To reduce the complexity of the collaborative development process, many of the repetitive human activities that are part of the development workflow are being automated by CI/CD tools that help to increase the productivity and quality of software projects. Social coding platforms aim to integrate all this tooling and workflow automation in a single encompassing environment. These social coding platforms gave rise to the emergence of development bots, facilitating the integration with external CI/CD tools and enabling the automation of many other development-related tasks.
\github, the most popular social coding platform, has introduced \actions to automate workflows in its hosted software development repositories since November 2019. This chapter explores the ecosystems of development bots and \actions and their interconnection. It provides an extensive survey of the state-of-the-art in this domain, discusses the opportunities and threats that these ecosystems entail, and reports on the challenges and future perspectives for researchers as well as software practitioners.\\
{\color{blue}This is a preprint of the following chapter: Mairieli Wessel, Tom Mens, Alexandre Decan, Pooya Rostami Mazrae,``The \github Development Workflow Automation Ecosystems'', published in ``Software Ecosystems: Tooling and Analytics'', edited by Tom Mens, Coen De Roover, Anthony Cleve, 2023, Springer Nature reproduced with permission of Springer Nature. The final authenticated version is available online at: http://dx.doi.org/[insert DOI]}
%
}
\newpage


\section{Introduction}
\label{WFA:sec:intro}

This introductory section presents the necessary context to set the scene. We start by introducing collaborative software development and social coding (\sect{WFA}{collab}). Next, we report on the emergence and dominance of \github as the most popular social coding platform (\sect{WFA}{github}). We continue with a discussion of the practices of continuous integration, deployment and delivery (\sect{WFA}{CI/CD}). Finally, we explain the workflow automation solutions of development bots and \actions that have emerged as highly interconnected ecosystems to support these practices, and that have become omnipresent in \github (\sect{WFA}{wf-ecosystem}).

We argue that these workflow automation solutions in \github constitute novel software ecosystems that are worthy of being studied in their own right.
More specifically, \sect{WFA}{bots} focuses on how development bots should be considered as an integral and important part of the fabric of \github's socio-technical ecosystem.
\sect{WFA}{GHA} focuses on \actions, and how this forms an automation workflow dependency network bearing many similarities with the ones that have been studied abundantly for packaging ecosystems of reusable software libraries.
\sect{WFA}{discussion} wraps up with a discussion about how both types of automation solutions are interrelated and how they are drastically changing the larger \github ecosystem of which they are part.

\subsection{Collaborative Software Development and Social Coding}
\label{WFA:sec:collab}

The large majority of today's software is either open source or depends on it to a large extent.
In response to a demand for higher-quality software products and faster time-to-market, open source software (OSS) development has become a continuous, highly distributed and collaborative endeavour~\cite{costa2011scale}. In such a setting,
development teams often collaborate on these projects without geographical boundaries \cite{herbsleb2007globalsweng}. It is no longer expected for software projects to have all their developers working in the same location during the same office hours. To achieve this new way of software development, specific collaboration mechanisms have been devised such as issue and bug tracking, pull-based development \cite{Gousios2014}, code reviews, commenting mechanisms and the use of social communication channels to interact with other project contributors.
Collaboration extends distributed software development from a primarily technical activity to an increasingly social phenomenon \cite{Tsay2014}. Social activities play an essential role in collaborative development, and become sometimes as critical as technical activities. They also come with their own challenges, for example because of cultural differences, language barriers or social conflicts \cite{holmstrom2006globaldev,Catolino2021}.

A multitude and variety of development-related activities need to be carried out during collaborative software development: developing, debugging, testing and reviewing code; quality and security analysis; packaging, releasing and deploying software distributions; and so on. This  makes it increasingly challenging for contributor communities to keep up with the rapid pace of producing and maintaining high-quality software releases.
It requires the orchestrated use of a wide range of tools such as version control systems, software distribution managers, bug and issue trackers, vulnerability and dependency analysers.

These tools therefore tend to be integrated into so-called \emph{social coding platforms} (\eg \gitlab, \github, BitBucket) that have revolutionised collaborative software development practices in the last decade because they provide a high degree of social transparency to all aspects of the development process \cite{Dabbish2012}. Social coding platforms aim to reconcile the technical and social aspects of software development in a single environment. It offers the project contributors a seamless interface and experience to contribute with their peers in an open and fully transparent workflow, where users can contribute bugs and feature requests through an issue tracking system, external contributors can propose code changes through a pull request mechanism, core software developers can push (\ie commit) their own code changes directly and accept and integrate the changes proposed by external contributors, and code review mechanisms allow code changes to be reviewed by other developers before they can be accepted \cite{Gousios2014}.

\subsection{The \github Social Coding Platform}
\label{WFA:sec:github}

\github has revolutionised software development since it was the first platform to propose a \emph{pull-based software development} process \cite{Gousios2014, Arora2016}.
The pull-based model allows to make a distinction between direct contributions from a typically small group of core developers with commit access to the main code repository, and indirect contributions from external contributors that do not have direct commit access.
This allows external contributors to propose code changes and code additions through so-called \emph{pull requests} (PR). To do so, these contributors have to \emph{fork} the main repository, update their local copies with code changes, and submit PRs to request to pull these changes into the main code repository \cite{Gousios2016}. This indirect contribution method enables the project's maintainers to review the code submitted through each PR, test it, request changes to the submitter of the PR if needed, and finally integrate the PR into the codebase without getting involved in code development \cite{Gousios2016}.
A pull-based development process also comes at a certain cost, since it raises the need for integrators --~specialised project members responsible for managing others' contributions who act as guardians of the projects' quality~\cite{Gousios2015}.

The focus of this chapter will be on \github, since it is the largest and most popular social coding platform by far, especially for open source projects, and as a consequence it has been the focus of a significant amount of empirical research. It is a web-based platform on the cloud, based on the git version control system, that hosts the development history of millions of collaborative software repositories, and accommodating over 94 million users in 2022~\cite{octoverse2022}.

\github continues to include more and more support for collaborative software development such as a web-based interface on top of the git version control system, an issue tracker, the ability to manage project collaborators, the ability to have a discussion forum for each git repository, an easy way to manage PRs or even to submit new PRs directly from within the \github interface, a mechanism to create project releases, the ability to create and host project websites, the ability to plan and track projects, support for analysing outdated dependencies and security vulnerabilities, and metrics and visualisations that provide insights in how the project and its community is evolving over time.
\github also comes with a REST and GraphQL API to query and retrieve data from GitHub or to integrate GitHub repositories with external tools.
By late 2022, \github added a range of new features including: (i) github.dev, a web-based code editor that runs entirely in the internet browser to navigate, edit and commit code changes directly from within the browser; (ii) \github CodeSpaces, a more complete development environment that is hosted in the cloud; (iii) \github Packages to create, publish, view and install new packages directly from one's code repository; (iv) \github CoPilot, an AI-based tool that provides smart code auto-completion; and (v) GitHub Actions, a workflow automation tool fully integrated into GitHub.

\subsection{Continuous Integration and Deployment}
\label{WFA:sec:CI/CD}

Continuous integration (CI), deployment and delivery (CD) have become the cornerstone of collaborative software development practices.
CI practices were introduced in the late 90s in the context of agile development and extreme programming methodologies.
According to the agile manifesto principles ``our highest priority is to satisfy the customer through early and continuous delivery of valuable software''~\cite{beck2001manifesto}.
In their seminal blog~\cite{Fowler2000}, Fowler and Foemmel presented CI as a way to increase the speed of software development while at the same time improving software quality and reducing the cost and risk of work integration among distributed teams.
They outlined core CI practices to do so, including frequent code commits, automated tests that run several times a day, frequent and fully reproducible builds, immediately fixing broken builds, and so on.
CD practices, on the other hand, aim at automating the delivery and deployment of software products, following any changes to their code~\cite{Chen2021GHA}.
Key elements of continuous deployment are the creation of feasible, small, and isolated software updates, that are automatically deployed immediately after completion of the development and testing~\cite{savor2016continuous}.

Many self-hosted CI/CD tools and cloud-based CI/CD services automate the integration of code changes from multiple contributors into a centralised repository where automated builds, tests, quality checks and deployments are run. Popular examples of such CI/CD solutions are Jenkins, Travis, CircleCI and Azure DevOps.
They have been the subject of much empirical research over the last decades. An excellent starting point is the systematic literature review by Soares et al. \cite{soares2022effects}, covering 106 research publications reporting on the use of CI/CD. This review aimed at identifying and interpreting empirical evidence regarding how CI/CD impacts software development.
It revealed that CI/CD has many benefits for software projects. Besides the aforementioned cost reduction, quality and productivity improvement, it also comes with a reduction of security risks, increased project transparency and predictability, greater confidence in the software product, easiness to locate and fix bugs and improved team communication.
CI can also benefit pull-based development by improving and accelerating the integration process.

CI/CD services have also been built into social coding platforms.
With GitLab CI/CD, \gitlab has already featured CI/CD capabilities since November 2012. BitBucket has supported Pipelines since May 2016. Based on popular demand, in response to this support for CI/CD in competing social coding platforms, \github officially began supporting CI/CD through \gha in August 2019, and the product was released publicly in November 2019.
Before the release of GitHub Actions, Travis used to be the most popular CI/CD cloud service for GitHub repositories~\cite{beller2017oops}.
However, quantitative evidence has revealed that Travis is getting replaced by \gha at a rapid pace~\cite{Golzadeh2021SANER}. 
Additional qualitative evidence has revealed the reasons behind this replacement and the added value that \gha is bringing in comparison to Travis~\cite{Rostami2023CI}.

\subsection{The Workflow Automation Ecosystems of GitHub}
\label{WFA:sec:wf-ecosystem}

The previous sections have highlighted that global software development, especially for OSS projects, is a continuous, highly distributed and collaborative endeavour.
The diversity in skills and interests of the projects' contributors, and the wide diversity of activities that need to be supported (e.g. coding, debugging, testing, documenting, packaging, deploying, quality analysis, security analysis and dependency analysis) make it very challenging for project communities to keep up with the rapid pace of producing and maintaining high-quality software releases.

Solutions to automate part of the software development workflow, such as the aforementioned CI/CD tools and services, have been successfully used to reduce this maintenance burden (see Section~\ref{WFA:sec:CI/CD}).
However, these tools do not support the entire range of project-related activities for which automation could come to the rescue.
There are many repetitive and time-consuming social and technical activities for which, traditionally, CI/CD tools did not provide any support. Some examples of these are
welcoming newcomers, keeping dependencies up-to-date, detecting and resolving security vulnerabilities, triaging issues, closing stale issues, finding and assigning code reviewers, encouraging contributors to remain active, and software licencing.
To help project contributors in carrying out these activities, CI/CD solutions have been complemented by novel workflow automation solutions:

\bigskip\noindent\textbf{Development bots.} A well-known and very popular example of such workflow automation solutions is what we will refer to as \emph{development bots}. Erlenhov \etal~\cite{Erlenhov2019} consider development bots to be artificial software developers who are autonomous, adaptive, and have technical as well as social competence.
Such automated software development agents have become a widely accepted interface for interacting with human contributors and automating some of their tasks.
A study by Wang \etal revealed that bots are frequently used in the most popular OSS projects on \github~\cite{Wang2022-butler}.
These bots tend to be specialised in specific activities, belonging to the following main categories:
CI/CD assistance, issue and PR management, code review support, dependency and security management, community support, and documentation generation.
Section~\ref{WFA:sec:bots} will discuss in detail how such bots are used on \github to automate part of the software development workflow, and how these bots form an integral part of the socio-technical ecosystem of software contributors and software projects.
More specifically, bots affect the social interaction within a software project, as they influence how human contributors communicate and collaborate, and may even change the collaboration patterns, habits and productivity of project contributors~\cite{Lebeuf2017a}.

\bigskip\noindent\textbf{\gha.}
Another popular mechanism to automate development activities in \github repositories is using \emph{\gha}, a workflow automation service officially released in November 2019. Its deep integration into \github implies that \gha can be used not only for automating  traditional CI/CD services such as executing test suites or deploying new releases, but also to facilitate other activities such as code reviews, communicating with developers, and monitoring and fixing dependencies and security vulnerabilities.
\gha allows project maintainers to define automated workflows for such activities. These workflows can be triggered in a variety of ways such as commits, issues, pull requests, comments, schedules, and many more~\cite{Chandrasekara2021}.
\gha also promotes the use and sharing of reusable components, called \emph{Actions}, in workflows. These Actions are distributed in public \github repositories and on the \github Marketplace.\footnote{\url{https://github.com/marketplace?type=actions}}
They allow developers to automate their workflows by easily integrating specific tasks (e.g., set up a specific programming language environment, publish a release on a package registry, run tests and check code quality) without having to write the corresponding code.
Only 18 months after its introduction, \gha has become the most dominant CI/CD service on \github~\cite{Golzadeh2021SANER}.
Section~\ref{WFA:sec:GHA} presents this ecosystem of reusable Actions in more detail.
This ecosystem forms a technical dependency network that bears many similarities with traditional package dependency networks of reusable software libraries (such as npm for JavaScript, RubyGems for Ruby, NuGet for .NET, Packagist for PHP, CRAN for R, Maven for Java) that have been the subject of many past empirical studies (\eg~\cite{decan:emse:2019}).

\medskip
The two aforementioned workflow automation solutions are increasingly used in OSS projects on \github, partly because of their tight integration into the social coding platform, thereby effectively transforming the software development automation landscape.
It therefore seems fair to claim that they form new \emph{development workflow automation ecosystems} that are worthy of being investigated in their own right.
Research on these ecosystems is still in its infancy, given the relative novelty of the proposed automation solutions.
Development bots and workflows that rely on \gha are already used in hundreds of thousands of \github repositories, and their usage continues to increase (the Marketplace of \gha has been growing exponentially since its introduction), justifying the need for further studies on the evolution of these ecosystem and their impact on collaborative software development practices.

\section{Workflow Automation Through Development Bots}
\label{WFA:sec:bots}

As explained in Section~\ref{WFA:sec:wf-ecosystem},
development bots emerged on social coding platforms such as \github to enable the automation of various routines and time-consuming tasks previously assigned only to human developers.
This section explores how bots are an integral part of \github's socio-technical collaborative development ecosystem. Considering the workflow automation provided by development bots, we focus on the various usage scenarios, advantages, shortcomings, challenges and opportunities of using them.

\subsection{What Are Development Bots?}

Development bots that reside on social coding platforms such as \github are often seen as workflow automation providers, due to their ability to react to certain stimuli, such as events triggered by human developers or other tools, and automate routine development-related tasks in response. To a certain extent, bots may act autonomously~\cite{Wyrich2019,Erlenhov2019}. In open source repositories, bots can leverage the public availability of software assets, including source code, discussions, issues and comments.

Besides automatically executing activities, development bots may also exhibit human-like traits. 
Erlenhov \etal~\cite{Erlenhov2019} describe bots based on their social competence, which varies from very simple identity characteristics (e.g., a human-like name or profile picture) to more sophisticated ones such as artificial intelligence and the ability to adapt to distinct scenarios.
In practice, bots that are active in \github repositories are automated agents that interact with the \github platform in essentially the same way as a typical human developer would be expected to: they possess a \github account, commit code, open or close issues or PRs, and comment on all of the above.
Some bots have an official integration with \github and are publicly available as Apps in the \github marketplace.\footnote{\url{https://github.com/marketplace?type=apps}}
These \emph{official} bots are properly tagged as such in the various activities they make in the \github platform.

Bots can also be used as an interface between human developers and other software services, such as external CI/CD tools or other third-party applications. Such bots provide additional value on top of the services they offer an interface for, by providing new forms of interaction with these services, or by combining multiple services.

\begin{figure}[!h]
  \centering
  \includegraphics[width=0.85\columnwidth]{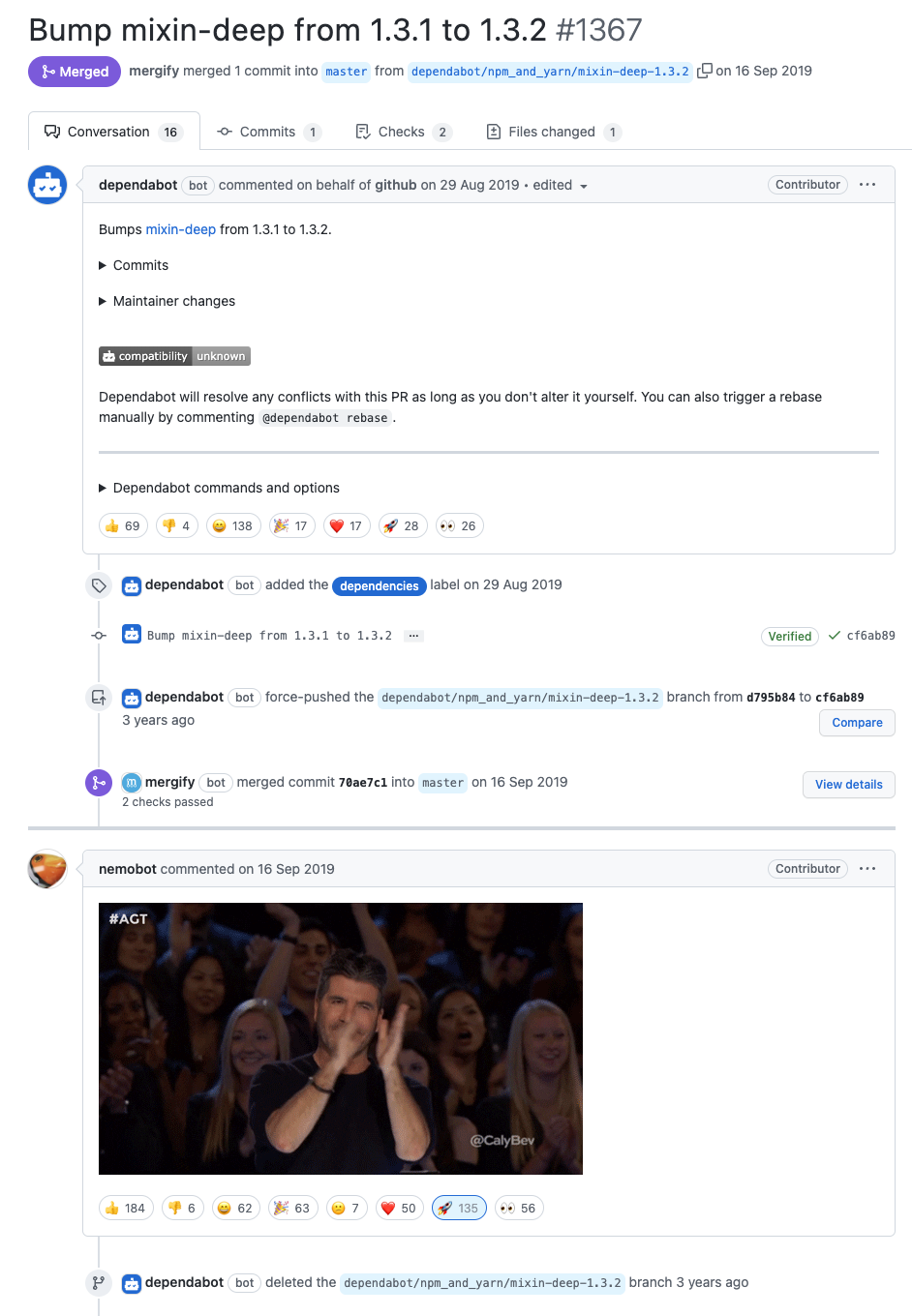}
  \caption{Example of multiple bots interacting within the same PR.}
  \label{fig:botInteraction}
\end{figure}

One particularly interesting example is \texttt{Dependabot}, a dependency management tool responsible for creating PRs in \github repositories to propose to upgrade dependencies, in order to resolve or reduce the risk of security vulnerabilities or bugs. \texttt{Dependabot} acts as an interface between the project maintainer, who is responsible for keeping the project dependencies up to date, and the package managers (such as \textsf{npm} for JavaScript) that expose the reusable packages that the project depends on. While originally it used to be a third-party service, \texttt{Dependabot} is now deeply integrated into the \github platform and has become one of the most popular dependency management bots, accounting for more than 7.7 million dependency updates in OSS projects~\cite{wyrich2021bots}. 
A well-known alternative is \texttt{renovatebot}.\footnote{\url{https://github.com/renovatebot/renovate}}

Bots can even create, review, and decide whether to integrate the changes made in a PR into the repository by themselves in complete autonomy.  Figure~\ref{fig:botInteraction} provides an example of multiple bots interacting as part of a single PR.
There is not a single human contributor involved in this interaction. The PR is triggered by a recommendation by \texttt{Dependabot} to update a dependency. The \texttt{mergify} bot reacts to this by verifying if the proposed change passes all checks, and accepts and merges the PR. Finally, \texttt{nemobot} reacts with a visual comment applauding the merged PR.

From a research viewpoint, the increasing use of bots raises the need for large-scale empirical studies on bot usage in social coding platforms such as \github.
Such studies enable us to assess whether bots serve their intended purpose, and whether their introduction has any positive or negative side-effects on the socio-technical fabric of the project or ecosystem in which they are used.
To enable such empirical studies, it is necessary to determine which projects rely on bots, and which user accounts actually correspond to bots. Several bot detection heuristics have been proposed to automatically identify bot contributions~\cite{Bodegha2021,AbdellatifBotHunter2022,BIMAN}. 
BIMAN \cite{BIMAN} relies on bot naming conventions, repetitiveness in commit messages, and features related to files changed in commits. BoDeGHa \cite{Bodegha2021} relies on comment patterns in issue and PR comments in GitHub repositories, based on the assumption that bots tend to use different and fewer comment patterns than humans.
BotHunter \cite{AbdellatifBotHunter2022} additionally relies on features corresponding to profile information (\eg account name) and account activity (\eg median daily activity) to identify bot accounts more accurately.
BoDeGiC \cite{Bodegic} allows to detect bots in git repositories based on commit messages, and has been trained using the classification model of BoDeGHa.

An important challenge when identifying automated contributions by bots is the presence of so-called \emph{mixed accounts} -- accounts used by a human developer and a bot in parallel -- exhibiting both human-like and bot-like behaviour~\cite{Bodegha2021}. 
Not properly detecting such cases is likely to lead to false positives and false negatives during bot detection, which may affect the outcome of empirical analyses.
Cassee \etal~\cite{CasseeMixedAccounts} have shown that existing classification models are not suitable to reliably detect mixed accounts.

\subsection{The Role of Bots in \github's Socio-technical Ecosystem}

An important characteristic of bots is that they form an integral part of \github's socio-technical ecosystem of collaborative software development. To consider them as such, we adopt an \emph{ecosystemic} and \emph{socio-technical} viewpoint, similar to Constantinou and Mens \cite{Constantinou2017} who viewed a software ecosystem as a socio-technical network that is composed of a combination of technical components (e.g., software projects and their source code history) and social components (e.g., contributor and communities involved in the development and maintenance of the software). 

An interesting novelty of bots is that, while they are technical components themselves (since they are executable software artifacts), they should also be considered as being social components, since they play a crucial role in the social aspects of the ecosystem. The assistance provided by bots, as new voices in the development conversation~\cite{monperrus2019explainable}, has the potential to smooth and improve the efficiency of developers' communication. Wessel \etal~\cite{wessel2022emse} have shown that the number of human comments decreases when using bots, which usually implies that the number of trivial discussions decreases. Indeed, bots are meant to relieve, augment and support the collaborative software development activities that are carried out by the human contributors that jointly develop and maintain large software projects. Moreover, bots often interact with human collaborators (and with other bots) using the same interface as humans do.

\begin{figure}[!h]
  \centering
    \includegraphics[width=0.8\columnwidth]{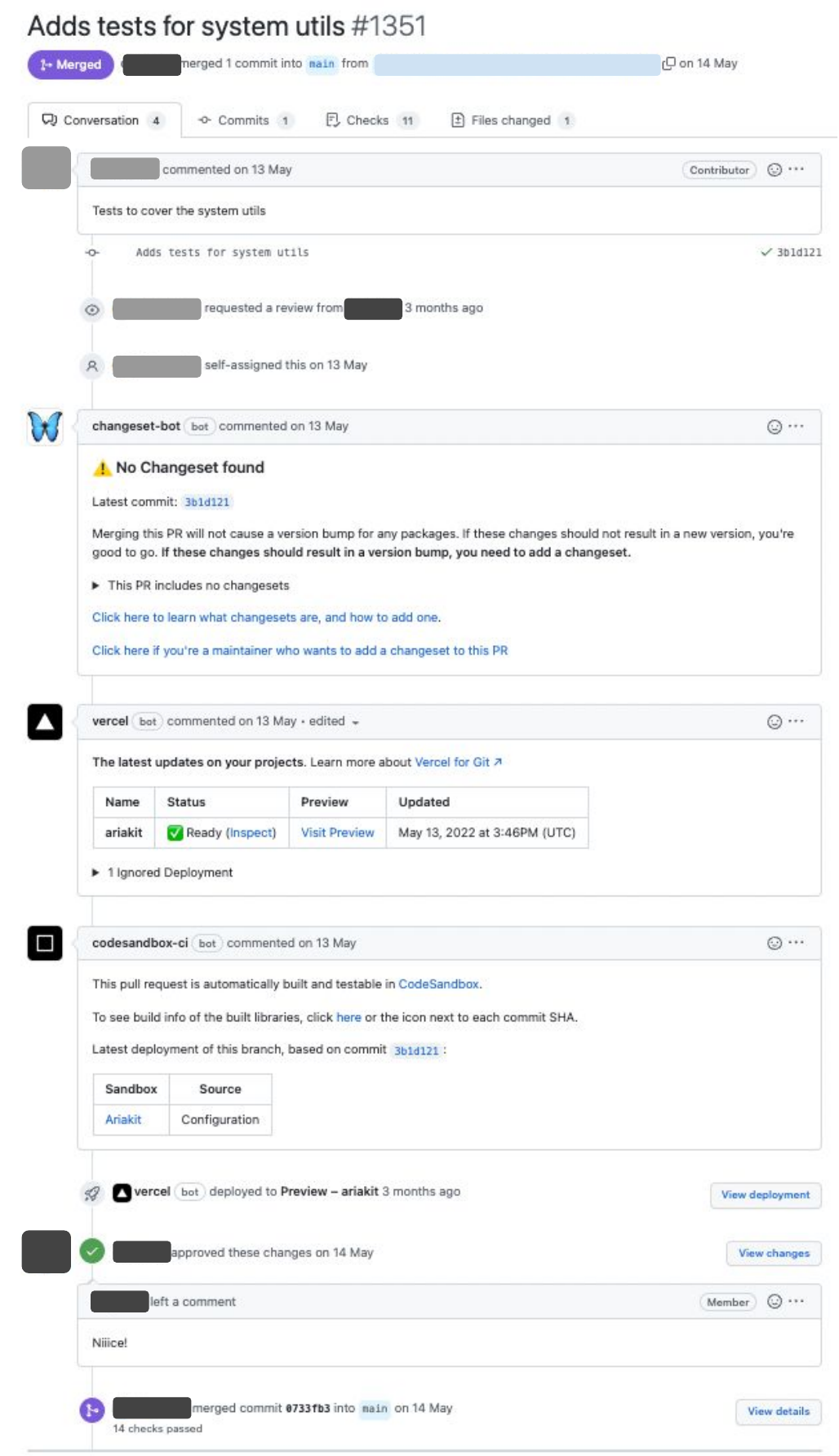}
  \caption{Example of an interaction between two humans and three bots within a single PR.}
  \label{fig:botHumanInteraction}
\end{figure}

Figure~\ref{fig:botHumanInteraction} illustrates an exemplary case of the role that bots play in this socio-technical ecosystem. A human contributor submits a PR to add tests to a particular project module. The first bot to react to the PR, \texttt{changeset-bot}, verifies whether the changeset file was updated and the proposed change will be released into a specific version of the packages implemented in the repository.
Then, \texttt{vercel} bot deploys the code to the third-party application Vercel and provides a URL for the developers to inspect a  deployment preview in the PR.
Next, the \texttt{codesandbox-ci} bot provides the URL of an isolated test environment to validate the changes made in the PR. Finally, the human project maintainer approves the changes, reacts with a comment, and merges the PR.

Like the many roles human software developers can fulfil, a variety of bots have become highly active actors in every phase of the development automation workflow.
Thanks to the continuous efforts of practitioners and researchers, a wide range of development bots are available for use by developers~\cite{Lebeuf2017a,Wang2022-butler}.
Wang~\etal\cite{Wang2022-butler} have shown that bot usage is common practice in OSS development. 
Through repository mining of 613 \github repositories, they found 201 different bots. Similar to prior research by Wessel \etal~\cite{power_of_bots}, the authors provided a classification of bots according to their main role in the repository. These categories include CI/CD assistance, issue and PR management, code review support, dependency and security management, community support, and documentation generation.

In addition to the aforementioned examples of bots, other sophisticated bots have been proposed in the literature.
Wyrich and Bogner~\cite{Wyrich2019}, for example, proposed a bot that automatically refactors the source code of a project. 
Their goal was to eliminate the need for developers to manually find and correct code smells, as this task can be very time consuming and may require certain expertise. Therefore, the bot was designed to act autonomously, integrating into the natural workflow of the development team on \github. 
The bot makes code changes corresponding to proposed code refactorings, and submits a PR with these changes. Project maintainers can review these changes and decide to integrate them into the code base.

\subsection{Advantages of Using Development Bots}

Development bots generally execute tasks that would otherwise have to be performed manually by humans. Through interviews with industry practitioners, Erlenhov \etal~\cite{erlenhov2020empirical} found that bots are used either because they improve productivity or enable activities that are not realistically feasible for humans~\cite{erlenhov2020empirical}. Some software practitioners stress that bots are able to carry out certain tasks better than humans due to their availability, scalability, and capacity to process large amounts of data~\cite{erlenhov2020empirical}. For example, bots can handle tasks continuously 24/7 without ever needing to take a break. Song and Chaparro \cite{SongBee} designed \texttt{BEE}, a bot that automatically analyses incoming issues on GitHub repositories and immediately provides feedback on them. Due to \texttt{BEE}'s prompt reaction, issue reporters can more quickly gain a general idea of what is missing without waiting for the project maintainers' feedback. Bots also scale, increase consistency, and mitigate human errors.
In terms of productivity increase, bot usage is frequently motivated by the necessity of spending less time on routine, time-consuming, or tedious tasks. Automating such activities through bots allow developers to focus on their core code development and review tasks~\cite{Storey2016,erlenhov2020empirical}. Mirhosseini and Parnin~\cite{mirhosseini2017can} analysed automated PRs created by \texttt{greenkeeper}, a bot to update dependencies, similar to the ones created by \texttt{dependabot}. Such a bot avoids manually monitoring for new releases in the packages. The results show that OSS repositories that use the bot upgraded the dependencies 1.6 times more regularly than repositories that did not use any other bots or tools.

Specifically in the context of code reviews, Wessel \etal~\cite{wessel2020expect} carried out a survey with 127 software project developers to investigate the advantages of adopting bots to support code review activities. Their study confirmed the results of Erlenhov \etal~\cite{erlenhov2020empirical}. The main reasons for adopting bots are related to improving developer feedback, automating routine tasks, and ensuring high-quality standards. Interestingly, developers also report benefits related to interpersonal relationships. According to the surveyed developers, negative feedback in an automatic bot report feels less rude or intimidating than if a human would provide the same feedback. They also report that by providing quick and constant feedback, bots reduce the chance that a PR gets abandoned by its author.

Bots can also help to support developers unfamiliar with a software project or with specific software engineering practices and technologies. For example, Brown and Parnin~\cite{brown2021nudging} propose a bot to nudge students toward applying better software engineering practices. They designed a bot that provides daily updates on software development processes based on students' code contributions on \github. They show that such a bot can improve development practices and increase code quality and productivity.

The use of bots to automate development workflows can also result in a change in the habits of project contributors. Wessel \etal~\cite{wessel2020effects} investigated how activity traces change after the adoption of bots. They observed that, after bot adoption, projects have more merged PRs, fewer comments, fewer rejected PRs, and faster PR rejections. Developers explain that some of these observed effects are caused by increased visibility of code quality metrics, immediate feedback, test automation, the increased confidence in the process, change in the discussion focus, and the fact that bot feedback pushes contributors to take action. 

In summary, the literature suggests that developers who employ bots primarily expect \textbf{improved productivity}~\cite{erlenhov2020empirical,wessel2020expect}. This, however, surfaces in different ways depending on the context and the tasks the bot performs. 
\textbf{Automating time-consuming or tedious tasks} and \textbf{collecting dispersed information} (i.e., information gathering) are some ways to improve productivity.
Developers also emphasize that bots may \textbf{perform some tasks better than humans} (e.g., handling tasks 24/7 and at scale, increasing consistency, and mitigating human error).

\subsection{Challenges of Using Development Bots}

Despite the numerous benefits leveraged by using development bots, several challenges have been reported concerning the workflow automation provided by them~\cite{bot_influence,erlenhov2020empirical}.
Some bots have been studied in detail, revealing the challenges and limitations of their PR interventions~\cite{Brown2019,mirhosseini2017can,Peng2019}.

\medskip \noindent \textbf{Trust.} Trusting a bot to act appropriately and reliably is challenging~\cite{erlenhov2020empirical}.
A side effect of overly relying on bots is that humans no longer question whether these bots are taking the correct actions since they assume bots to be experts in their tasks.
Therefore, developers can be caught off-guard by excessive incorrect outcomes from bots~\cite{erlenhov2020empirical}. 
A key solution to increase trust is building a reliable testing environment that allows developers to try out bots and avoid unanticipated problems.

\medskip \noindent \textbf{Discoverability and configuration.} To confirm the challenges caused by development bots in PR interactions, Wessel \etal~\cite{bot_influence} interviewed 21 practitioners. Their study revealed several challenges raised by bot usage, such as discoverability and configuration issues. 
Developers complained about the lack of contextualized actions, limited and burdensome configuration options, and technical overhead to host and deploy their own bot. Moreover, the overload of information generated by bots when interacting on PRs has appeared as the most prominent challenge.

\medskip \noindent \textbf{Interruption and noise.} Developers constantly struggle with interruptions and noise produced by bots~\cite{erlenhov2020empirical}.
For instance, Brown and Parnin~\cite{Brown2019} analysed \texttt{tool-recommender-bot}, a bot that
automatically configures a project to use an open source static analysis tool for Java code and then submits a PR with a generic message explaining how the proposed tool works.
They reported that this bot still needs to overcome problems such as notification workload. They applied \texttt{tool-recommender-bot} in real projects for evaluation purposes. Only two PRs out of 52 proposed recommendations were accepted. 
Peng and Ma~\cite{Peng2019} studied how developers perceive and work with \texttt{mention bot}, a reviewer recommendation bot created by Facebook. It automatically tags a potential reviewer for a PR depending on the files changed. Project maintainers with higher expertise (i.e., maintainers who contributed more frequently) in a particular file are more likely to be suggested as reviewers by the bot. 
The study found that \texttt{mention bot} reduced contributors' effort in identifying proper reviewers. As a negative side-effect, however, developers were bothered by frequent review notifications when dealing with a heavy workload.

Wessel \etal~\cite{bot_influence} introduced a theory about how certain bot behaviours can be perceived as noisy. 
Indeed, many bots provide several comments when an issue or PR is opened by a contributor, with dense information and frequently overusing visual elements. 
Similarly, bots perform repetitive actions such as creating numerous PRs (e.g., to update the many dependencies a project can have) and leaving dozens of comments in a row (e.g., to report on test coverage each time a new commit is added to the PR). 
These situations can lead to information and notification overload, disrupting developers' communication.
Oftentimes, the problem is not a singular bot that is too verbose, but a combination of multiple bots that are simultaneously active and, together, lead to information overload~\cite{erlenhov2020empirical}.

\begin{figure}[!ht]
    \centering
    \includegraphics[width=0.8\linewidth]{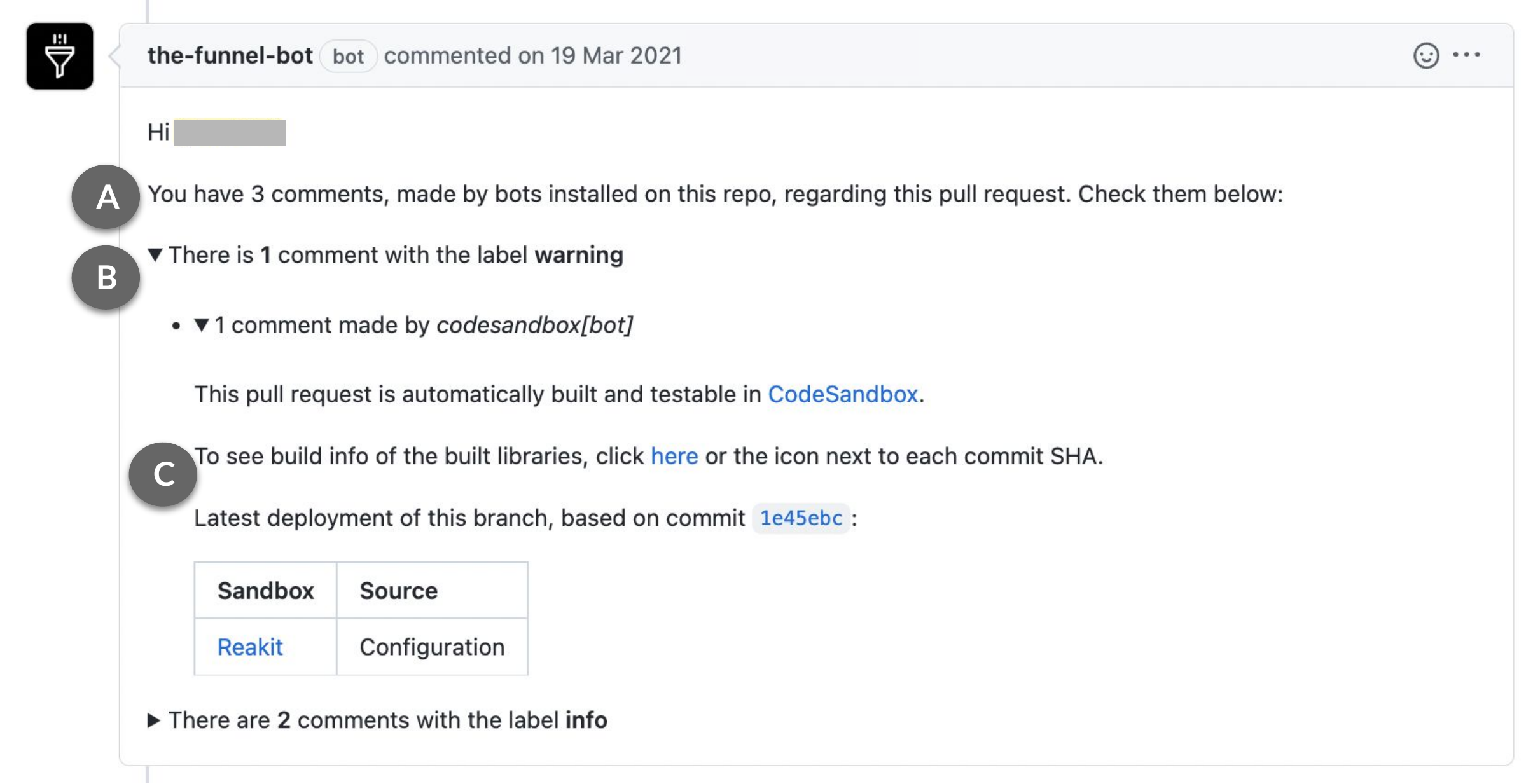}
    \caption{Example of PR comment created by \texttt{FunnelBot}.}
    \label{fig:metabot}
\end{figure}

Researchers have attempted to create solutions to reduce the information overload created by bots. Wessel \etal~\cite{WesselICSE2022} suggested creating better ways to represent the information of bots, such as clearer summaries of pull requests.  Ribeiro \etal~\cite{ribeiro2022} implemented \texttt{FunnelBot} that integrated these suggestions.
Figure \ref{fig:metabot} shows an example of a PR comment posted by \texttt{FunnelBot}. 
The comment shows (A) an introductory message, (B) a list with all groups of bot messages collapsed, and (C) one expanded example where we can see the \texttt{CodesandBox} comment.

\section{Workflow Automation Through \actions}
\label{WFA:sec:GHA}

As explained in Section~\ref{WFA:sec:wf-ecosystem}, software development workflows can be automated using different techniques, including CI/CD solutions (presented in Section~\ref{WFA:sec:CI/CD}) and development bots (presented in Section~\ref{WFA:sec:bots}).
The third way is \gha, which is the focus of the current section.
We explain what \actions are, how prevalent they are, and how they constitute an ecosystem of their own. We also discuss the potential challenges this novel ecosystem is confronted with.

\subsection{What is \actions?}

The \github social coding platform has introduced \actions as a way to enable the specification and execution of automated workflows. It started as a beta product in 2018 providing the possibility to create Actions inside containers to augment and connect software development workflows. When the product was officially released to the public in November 2019, \actions also integrated a fully-featured CI/CD service, answering the high demand of \github users to provide CI/CD support similar to what was already available in competing social coding platforms such as GitLab and Bitbucket~\cite{Dabbish2012}.

\begin{figure}[!h]
  \centering
  \includegraphics[width=0.8\textwidth]{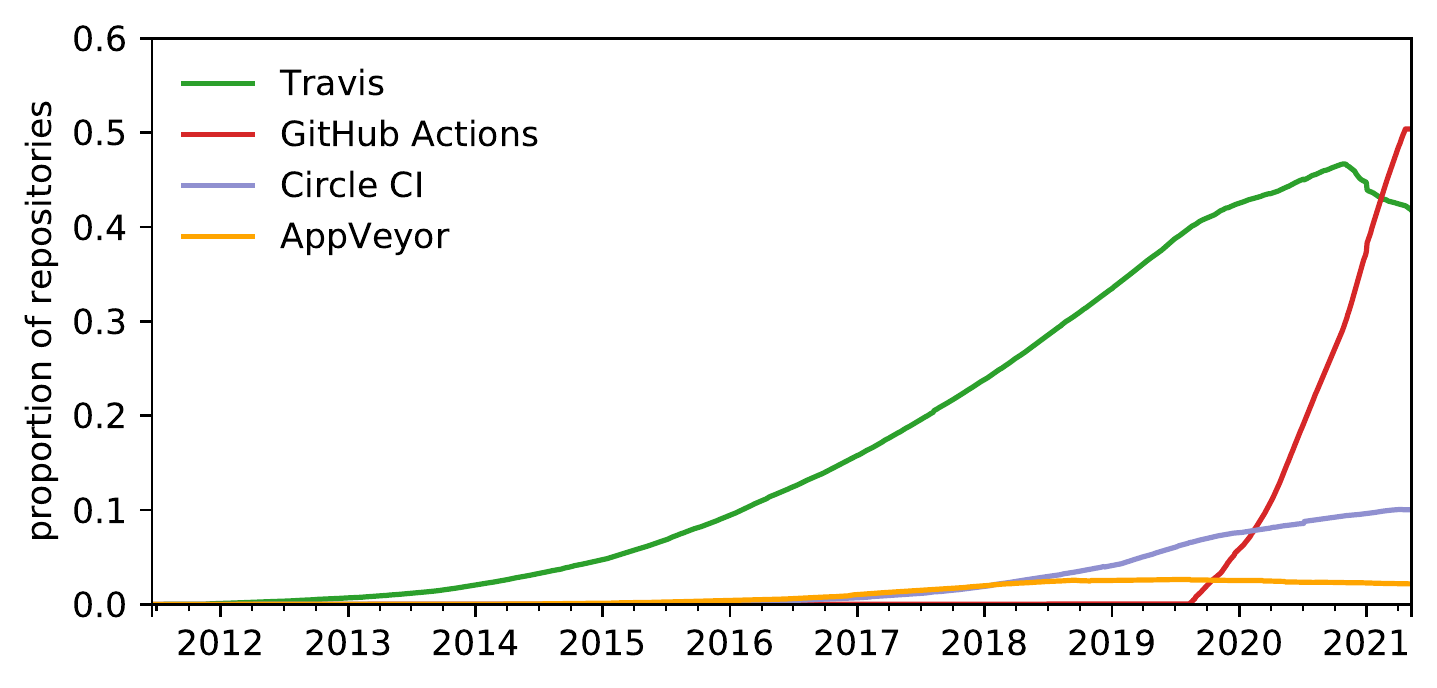}
  \caption{Evolution of the proportion of \github repositories using a specific CI/CD solution.}
  \label{fig:gha_usage_increase}

\end{figure}

Since its introduction, \actions has become the dominant CI/CD service on \github based on a quantitative study by Golzadeh \etal~\cite{Golzadeh2021SANER}, including more than $90K$ \github repositories. Figure~\ref{fig:gha_usage_increase} provides a historical overview of CI/CD usage in those repositories, starting from the first observation of Travis usage in June 2011. Initially, \github repositories primarily used Travis as a CI/CD service. Over time, other CI/CD solutions became used, but Travis remained the dominant CI/CD. When \actions entered the CI/CD landscape, it overtook the other CI/CD solutions in popularity in less than 18 months after its introduction.
Mazrae \etal \cite{Rostami2023CI} complemented this quantitative analysis by qualitative interviews to understand  the reasons behind \gha becoming the dominant CI/CD tool in \github, as well as why project maintainers decided to migrate to \gha primarily.
The main reported reasons were the seamless integration into \github, the ease of use, and great support for its reusable Actions.

\actions allows repository maintainers to automate a wide range of tasks. In addition to providing typical CI/CD services such as building code, executing test suites and deploying new releases, \actions' tight integration with \github enables it to include better support of third-party tools, build support for well-known operating systems and hardware architectures, and more scalable cloud-based hardware to produce results faster.
\actions also facilitate the communication between the project and external tools (such as third-party CI/CD services) and easier dependency and security monitoring and management~\cite{decanuse}.

\begin{figure}[!h]
  \centering
  \includegraphics[scale=0.7]{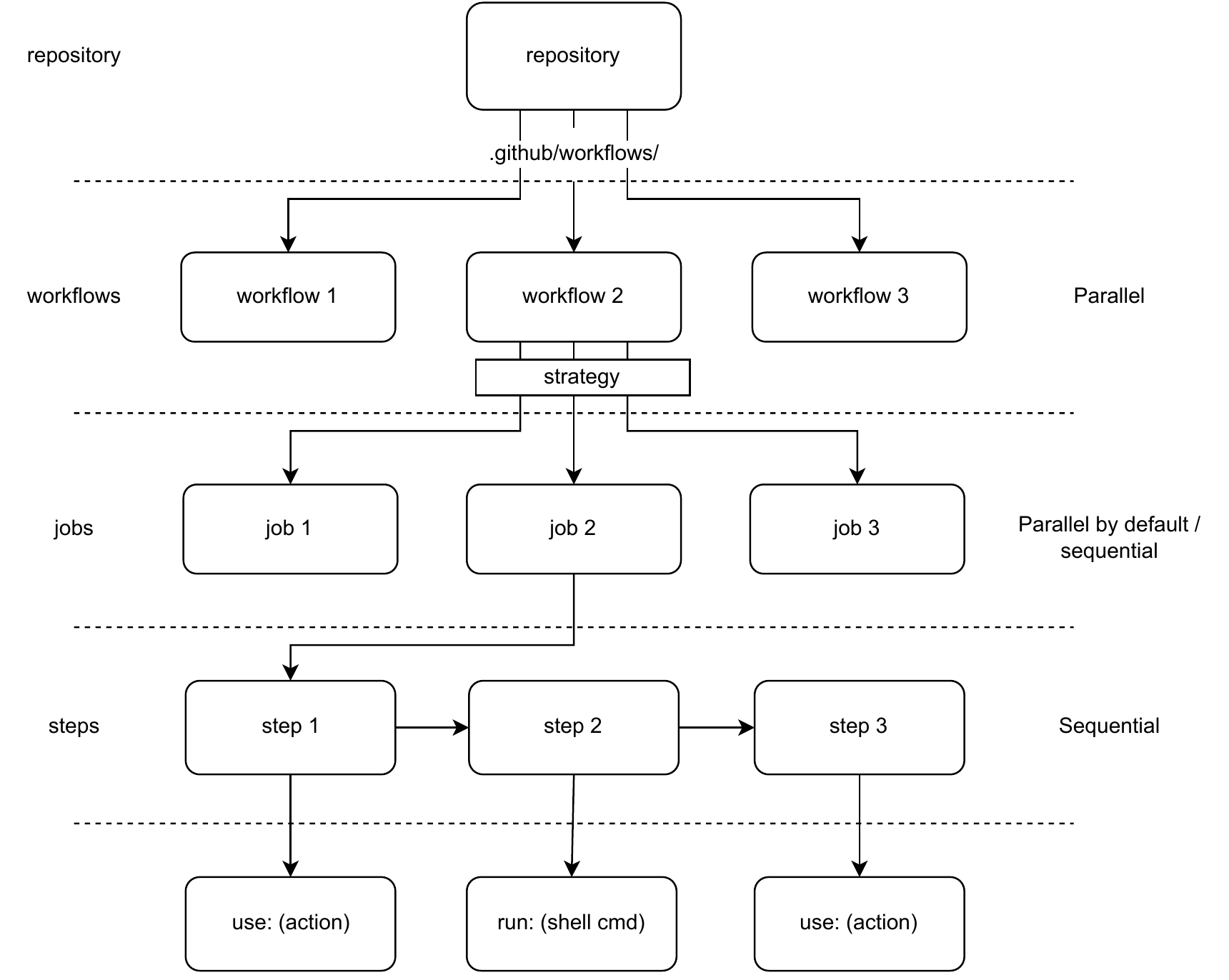}
  \caption{Schematic representation of the structure of a GitHub workflow  description.}
  \label{fig:actions_structure}
\end{figure}

\bigskip \noindent \textbf{Specifying executable workflows.}
\actions is based on the so-called concept of executable \emph{workflows} that can be defined by maintainers of \github repositories. The structure of a workflow is schematically presented in Figure~\ref{fig:actions_structure} and explained below.

A workflow constitutes a configurable automated process that is defined by a YAML file added to the \texttt{.github/workflows} directory of the \github repository.
A workflow can be executed based on events specified in the workflow description that act as a trigger for running the workflow. Examples of such triggers are commits, issues, PRs, comments, schedules, or even manual invocation~\cite{Chandrasekara2021}.
The example workflow in \lst{WFA}{YML_example} (lines 3 to 6) defines three possible triggers: upon committing (\texttt{push:}) or receiving a PR (\texttt{pull\_request:}), or based on a specified time schedule (\texttt{cron: "0 6 * * 1"}).

\begin{lstlisting}[float,language=yaml,frame=single,numbers=left,label={WFA:lst:YML_example},basicstyle=\footnotesize\ttfamily,caption={Example of a YAML workflow file.}]
name: Test project
on:
  push:
  pull_request:
  schedule:
    - cron: "0 6 * * 1"
  jobs:
    build-and-test:
      strategy:
        matrix:
          os: [ubuntu-22.04, windows-latest]
          python: ["3.6", "3.7", "3.8", "3.9", "3.10"]
      runs-on: ${{ matrix.os }}
      steps:
        - uses: actions/checkout@v2
        - name: Set up Python
          uses: actions/setup-python@v2
          with:
            python-version: ${{ matrix.python }}
        - name: Install dependencies
          run: |
            pip install -r requirements.txt
            pip install pytest
        - name: Execute tests
          run: pytest
\end{lstlisting}

A workflow typically runs one \emph{job} in some virtual environment that is created to execute the job (e.g., an instance of some specific version of Ubuntu, macOS, or Microsoft Windows).
A workflow can also execute multiple jobs, in parallel (by default) or sequentially.
Workflows can define a \emph{matrix strategy} to automatically create and run parallel jobs based on the combination of variable values defined by the matrix.
This is for example useful if one would like to build and test source code in multiple versions of a programming language and/or on multiple operating systems.
In the example of \lst{WFA}{YML_example}, the matrix strategy (lines 10-13) specifies that the job will be run on 5 different versions of Python for two different operating systems.

To run a workflow specified in a \github repository, developers can use the infrastructure provided by \github, or rely on self-hosted runners if more specific hardware or operating systems are needed.
Each job is composed of a series of \emph{steps} that specify the tasks to be executed sequentially by the job. These steps can be simple shell commands to be run within the virtual environment (such as lines 22-24 in \lst{WFA}{YML_example}). Alternatively, steps can use and execute predefined reusable \emph{Actions}, that will be discussed below.

\bigskip \noindent \textbf{Reusable Actions.} Actions provide a reuse mechanism for \github workflow maintainers to avoid reinventing the wheel when automating repetitive activities~\cite{Chen2021GHA}. Rather than manually defining the sequence of commands to execute as part of a step (such as lines 22 to 24 of \lst{WFA}{YML_example}), if suffices to use a specific (version of a) reusable Action. For example, line 16 of \lst{WFA}{YML_example} (re)uses version 2 of \texttt{actions/checkout}, and line 18 (re)uses version 2 of \texttt{actions/setup-python}.
Actions are themselves developed through \github repositories.\footnote{The \github repositories for the Actions reused in \lst{WFA}{YML_example} are \url{https://github.com/actions/checkout} and \url{https://github.com/actions/setup-python}}

Workflows can reuse any Action shared in a public repository. To facilitate finding such Actions, the \github \emph{Marketplace} provides an interface for providers to promote their Actions, and for consumers to easily search for suitable Actions.\footnote{See \url{https://github.com/marketplace}. In addition to Actions, the  marketplace also promotes Apps, which are applications that can contain multiple scripts or an entire application.}
The number of Actions promoted on the Marketplace has been growing exponentially.
By December 2022, the Marketplace listed over 16K reusable Actions falling under $19$ different categories.
These categories contain a wide diversity of Actions, covering tasks such as setting up a specific programming language environment, publishing a release on a package registry, running tests or checking the code quality~\cite{decanuse}.

\subsection{Empirical Studies on \actions}
\label{WFA:sec:action-studies}
Given that \actions were publicly introduced in 2019, and despite the fact that \actions has become the dominating CI/CD solution on \github (according to Golzadeh \etal~\cite{Golzadeh2021SANER}), very few empirical studies have focused on \actions at the time of writing this chapter.

An early quantitative study by Kinsman \etal~\cite{kinsman2021software} in 2021 reported that, in a dataset of 416,266 \github repositories, only as little as 3,190 repositories (\ie less than 1\%) had been using \actions.
In 2022, Wessel \etal~\cite{wessel2022github} studied a dataset composed of the 5,000 most-starred \github repositories, and observed that 1,489 projects (\ie 29.8\%) had been using \actions.
Also in 2022, Decan \etal~\cite{decanuse} reports on a dataset of 67,870 active \github repositories in which 29,778 repositories (\ie 43.9\%) had been using \actions. These quantitative results reveal that \actions are prevalent in software development repositories on \github.
To complement these quantitative findings, in 2023, Saroar and Nayebi~\cite{Saroar2023} carried out a survey with 90 GitHub developers about the best practices and perception in using and maintaining \actions.

Table \ref{table: most_used_act_lang} reports the top 6 programming languages that most frequently coincide with \actions usage according to Decan \etal~ \cite{decanuse}. They observed that some programming languages are more likely to coincide with \actions usage than others: TypeScript and Go have a higher proportion of repositories resorting to \actions usage (respectively 58.5\% and 57.2\%) compared to JavaScript (34.9\%).
It is worth noting that the percentages of repositories using \actions are reported with respect to the language itself. For example, the number of Python repositories using \actions  is 1.52 times higher (5,654) than the TypeScript repositories using \actions (3,722). One can also observe the number of repositories for a specific language and its proportion to all the repositories in the dataset. For example, the 13,542 JavaScript repositories correspond to 19.6\% of all the repositories in the dataset.

\begin{table}[t]
\begin{center}
\caption{Top 6 languages with the highest proportion of \github repositories using \actions according to \cite{decanuse}.}
\label{table: most_used_act_lang}
\begin{tabular}{p{3.5cm}  r  r}
\toprule
   & \multicolumn{2}{c}{\bf \github repositories} \\
 \bf programming language~ & \multicolumn{1}{c}{all repositories~} & \multicolumn{1}{c}{~using \actions} \\
 \midrule
 JavaScript & 13,542 (19.6\%) & 4,730 (34.9\%) \\
 Python & 12,319 (17.8\%) & 5,654 (45.9\%) \\
 TypeScript & 6,362 (9.2\%) & 3,722 (58.5\%) \\
 Java & 6,105 (8.8\%) & 2,390 (39.2\%) \\
 C++ & 5,701 (8.2\%) & 2,331 (40.9\%) \\
 Go & 4,988 (7.2\%) & 2,854 (57.2\%) \\
\bottomrule
\end{tabular}
\end{center}
\end{table}

The same study also analysed which event types are mostly used for triggering workflows, reporting that \texttt{push:} and \texttt{pull\_request:} are the most frequent events triggering workflows, both used by more than half (respectively 63.4\% and 56.3\%) of all considered \github repositories relying on workflows. This is not surprising since commits and PRs are the most important activities in collaborative coding on \github.
The most frequently used Action is \texttt{actions/checkout} (used by 35.5\% of all steps and 97.8\% of all repositories). Other frequently used Actions are related to the deployment of a specific programming language environment (e.g., \texttt{setup-node}, \texttt{setup-python}, \texttt{setup-java}). Overall, 24.2\% of all steps use an Action of the form \texttt{setup-*}.

Finally, they observe that it is common practice to depend on reusable Actions, given that nearly all repositories ($>99\%$) that use workflows have at least one step referring to an Action. More than half of the steps in all analysed workflows (51.1\%) use an Action. However, this reuse is concentrated towards a limited set of Actions.
For example, the Actions that are officially provided by GitHub (\ie those actions of the form \texttt{actions/*}) account for 71.7\% of all steps that reuse an Action.
In addition to this, the Actions being reused tend to be concentrated in a few categories. Table \ref{table: most_freq_act_cat} provides the top 5 categories of Actions used by \github repositories, as reported by two independently conducted empirical studies \cite{wessel2022github, decanuse}.
Most of the reused actions belong to the `Utilities' and `Continuous Integration' categories, followed by `Deployment'. This suggests that \actions is being used mostly to automate the same kinds of activities as what traditional CI/CD tools are being used for.

\begin{table}[h!]
\begin{center}
\caption{Top 5 most frequent Action categories according to \cite{wessel2022github, decanuse}.}
\label{table: most_freq_act_cat}

\begin{tabular}{l r r}
 \toprule
   \bf Action category & \bf \% reported by~\cite{wessel2022github}~ &  \bf \% reported by~\cite{decanuse}\\
 \midrule
  utility & 24.9\% &  23.9\%\\
  continuous integration & 24.7\% &  17.3\%\\
  deployment & 9.6\% &  7.2\%\\
  publishing & 8.4\% &  6.9\%\\
 code quality & 7.7\% &  6.1\%\\
\bottomrule
\end{tabular}
\end{center}
\end{table}

Wessel \etal~\cite{wessel2022github} statistically studied the impact of using workflows on different aspects of software development like PRs, commit frequency, and issue resolution efficiency.
By comparing the activities in projects using \actions, during one full year before the use of \actions in the project and one full year after its usage, they used the technique of regression discontinuity analysis to provide statistical evidence and showed that after adding \actions to projects, there tend to be fewer accepted PRs, with more discussion comments and fewer commits, which take more time to merge. On the other hand, there are more rejected PRs, which contain fewer comments and more commits.

Wessel \etal~\cite{wessel2022github} studied discussions between developers about the usage of \actions in their software projects.
Out of the 5,000 analysed \github repositories, only 897 (18\%) had the Discussions feature enabled at the time of data collection, and 830 of those (17\%) contained at least one discussion thread. Focusing on this subset of repositories, they filtered the discussions containing the string ``GitHub Actions'', resulting in 573 posts in 458 distinct threads of 148 repositories.
The most discussed material about \actions, found in 28.8\% of all considered posts, was \emph{the need for help} with \actions. This reveals that developers
actively sought to learn more about how to use workflows effectively. A second popular category of discussion in the context of \actions, found in 19.0\% of all considered posts, were \emph{error messages or debug messages}. Developers were trying to solve issues related to using workflows and applications invoked via these workflows, such as linters or code review bots.
A third popular category, accounting for 14.6\% of all considered posts, involved discussions around \emph{reusing Actions}. This is expected, given that Actions are a relatively new concept that many developers are not familiar with.

\subsection{The \actions Ecosystem}
\label{WFA:sec:action-ecosystem}

As mentioned in Section~\ref{WFA:sec:wf-ecosystem}, \actions is part of the larger workflow automation ecosystem of \github that also includes bots and CI/CD solutions for automating development workflows in collaborative software projects.
Decan \etal~\cite{decanuse} suggested that \gha can and should be considered as a new emerging ecosystem in its own right.
Indeed, the \gha technology exhibits many similarities with more traditional software packaging ecosystems such as npm (for JavaScript), Cargo (for Rust), Maven (for Java) or PyPI (for Python) to name but a few.
Just as software development repositories on \github tend to depend on external packages distributed through the above package managers -- mainly to avoid the effort-intensive and error-prone practices of copy-paste reuse -- the same is valid for development workflows.
Maintainers of \github repositories can specify their workflows to directly depend on reusable Actions.
As such, \gha forms a kind of \emph{dependency network} that bears many similarities with the ones of software packaging ecosystems~\cite{decan2017empirical}.
The parallel with packaging ecosystems is quite obvious: automated workflows, as software clients, express \emph{dependencies} towards Actions (being the equivalent of reusable \emph{packages}) that can exist in different \emph{versions} or \emph{releases}.
Section~\ref{WFA:sec:action-studies} reported on quantitative evidence that resorting to reusable Actions in workflows has become a common practice.

\begin{figure}[!h]
  \centering
  \includegraphics[width=0.75\textwidth]{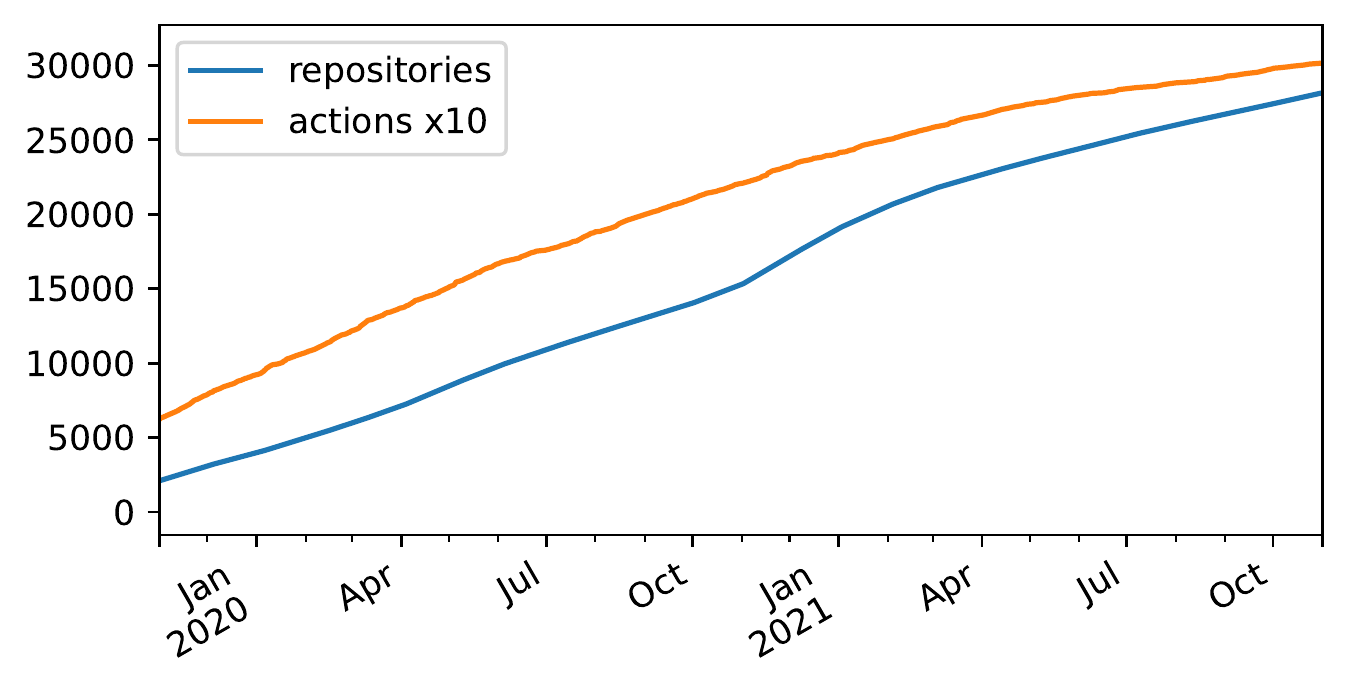}
  \caption{Evolution of the number of \github repositories using workflows (blue line) and the number of Actions used by these repositories (orange line, scaled by a factor 10 for ease of comparison).}
  \label{fig:evol}
\end{figure}

\smallskip \noindent \textbf{Continuous growth.}
Decan \etal~\cite{decan2017empirical} carried out a quantitative empirical analysis of the similarities and differences in the evolution of the dependency networks for seven different packaging ecosystems of varying sizes and ages, including Cargo, CPAN, CRAN, npm, NuGet, Packagist, and RubyGems.
They observed that these dependency networks tend to grow over time, both in size and in number of package updates.
While the vast majority of packages depend on other packages, only a small proportion of these packages account for most of the reuse (i.e., they are targeted by most of the reverse dependencies).
Decan \etal~\cite{decanuse} conducted a quantitative analysis of \gha and observed similar characteristics for the \gha dependency network: nearly all the repositories with \gha workflows depend on reusable Actions, and a limited number of Actions concentrate most of the reuse.
They analysed the evolution of the number of repositories using \gha workflows, and the number of Actions being used by these repositories. Figure~\ref{fig:evol} shows this evolution for the period 2020--2021, revealing a continuous growth of the \gha ecosystem, in terms of the number of \emph{consumers} (repositories using \gha workflows) as well as  \emph{producers} (Actions being reused by \github repositories).

\subsection{Challenges of the \actions Ecosystem}

While packaging ecosystems are extremely useful for their respective communities of software developers, they have  been shown to face numerous challenges related to dependency management~\cite{decan:emse:2019, kula2018developers, soto2021comprehensive}, outdatedness~\cite{decan2018evolution}, security vulnerabilities~\cite{decan2018impact,zimmermann2019small}, breaking changes~\cite{dietrich2019dependency,decan2021what}, deprecation~\cite{cogo2021deprecation} and abandonment of package maintainers~\cite{Constantinou2017,Avelino2019}.
We posit that \gha will suffer (and likely suffers already) from similar issues.

\medskip \noindent \textbf{Outdatedness.}
Software developers are continuously confronted with the difficult choice of whether, when, and how to keep their dependencies up to date.
On the one hand, updating a dependency to a more recent version enables them to benefit from the latest bug and security fixes. On the other hand, doing so exposes the dependent project to an increased risk of breaking changes, as well as to new bugs or security issues that may not even have been discovered yet.

The concept of {\em technical lag} was proposed to measure the extent to which a software project has outdated dependencies~\cite{gonzalez2017technical}. This lag can be quantified along different dimensions: as a function of time (how long has a dependency been outdated), version (how many versions is a dependency behind), stability (how many known bugs could have been fixed by updating the  dependency) and security (how many security vulnerabilities could have been addressed by updating the dependency).
Zerouali \etal~\cite{zerouali2019formal} formalized this concept in a measurement framework that can be applied at the level of packaging ecosystems. In particular, they analysed the technical lag of the npm packaging ecosystem, observing that around 26\% of the dependencies expressed by npm packages are outdated, and that half of these outdated dependencies target a version that is 270+ days older than the newer one. Other researchers have also applied technical lag to quantify outdatedness in software package dependency networks~\cite{decan2018evolution,Stringer2020}.
 The technical lag framework was also applied to the ecosystem of Docker containers distributed through Docker Hub~\cite{Zerouali2019Docker}.
\chap{IAC} provides more details on this matter.

In a similar vein, applying the technical lag framework to the \gha ecosystem would allow workflow developers to detect and quantify the presence of outdated Actions in workflows and help in updating them. It is important to do so since, despite the recency of \gha, according to Decan \etal~\cite{decanuse} at least 16\% of the dependencies in workflows are targeting an old major version of an Action.

\medskip \noindent \textbf{Adherence to semantic versioning.}
\emph{Semantic Versioning} (abbreviated to \emph{SemVer} hereafter) is another mechanism that has been proposed to assist software developers with the delicate trade-off between benefiting from security or bug fixes and being exposed to breaking changes in dependencies.
SemVer introduces a set of simple rules that suggest how to assign version numbers in packages to inform developers of dependent software about potentially breaking changes. In a nutshell, SemVer proposes a three-component version scheme \textsf{major.minor.patch} to specify the type of changes that have been made in a new package release.
Many software packaging ecosystems (such as npm, Cargo and Packagist) are mostly SemVer-compliant, in that most of their package producers adhere to the SemVer convention~\cite{decan2021what}.
Backward incompatible changes are signalled by an update of the \textsf{major} component, while supposedly compatible changes come with an update of either the \textsf{minor} or \textsf{patch} component.
This allows dependent packages to use so-called \emph{dependency constraints} to define the range of acceptable versions for a dependency (for example, it would be safe to accept all dependency updates within the same major version range if the dependency is trusted to be SemVer-compliant).

Maintainers of \gha workflows are exposed to a similar risk of incompatible changes in the Actions they use, whether these are logical changes (affecting the behaviour of the Actions) or structural changes (affecting the parameters or return values). Therefore, knowing whether an Action adheres to SemVer is helpful for maintainers of workflows depending on these actions, since they can assume minor and patch updates to be backward compatible and, therefore, free of breaking changes.
\github recommends reusing Actions in workflows by specifying only the major component of the Action's version, allowing workflow maintainers to receive critical fixes and security patches while maintaining compatibility. However, little is known about the actual versioning practices followed by producers and consumers of Actions.
Preliminary results suggest that GitHub's recommendation is widely followed since nearly 90\% of the version tags used to refer to an Action include only a major component~\cite{decanuse}.
However, unlike package managers, \gha offers no support for dependency constraints, implying that  producers of Actions are required to move these major version tags each time a new version of the Action is released.
Unless automated, this requirement introduces an additional burden on the Action producers~\cite{decanuse}, and calls for a more profound analysis of the kind of changes made in Action updates and of the versioning practices they follow.

\medskip \noindent \textbf{Security vulnerabilities.}
Another issue is that any software project is subject to security vulnerabilities.
Package dependency networks have made the attack surface of such vulnerabilities several orders of magnitude higher due to the widespread dependence on reusable software libraries that tend to have deep transitive dependency chains \cite{decan2018impact,zimmermann2019small,Alfadel2021,Dusing2021}.
For example, through a study of 2.8K vulnerabilities in the npm and RubyGems packaging ecosystems, Zerouali \etal~\cite{zerouali2022impact} found around 40\% of the packages to be exposed to a vulnerability due to their (direct or transitive) dependencies, and it often took months to fix them. They also observed that a single vulnerable package could expose up to two-thirds of all the packages depending on it.

We see no reason why the \gha ecosystem would be immune to this phenomenon. Indeed, relying on reusable Actions from third-party repositories or even from the Marketplace further increases the vulnerability attack surface.
Since a job in a workflow executes its commands within a runner shared with other jobs from the same workflow, individual jobs in the workflow can compromise other jobs they interact with.
For example, a job could query the environment variables used by a later job, write files to a shared directory that a later job processes, or even more directly interact with the Docker socket and inspect other running containers and execute commands in them.\footnote{\url{https://docs.github.com/en/actions/security-guides/security-hardening-for-github-actions\#using-third-party-actions}}
Multiple examples of security issues in workflows have been reported, sometimes with potentially disastrous consequences, such as manipulating pull requests to steal arbitrary secrets,\footnote{\url{https://blog.teddykatz.com/2021/03/17/github-actions-write-access.html}} injecting arbitrary code with workflow commands\footnote{\url{https://packetstormsecurity.com/files/159794/GitHub-Widespread-Injection.html}} or bypassing code reviews to push unreviewed code.\footnote{\url{https://medium.com/cider-sec/bypassing-required-reviews-6e1b29135cc7}}
Unfortunately, we are not aware of any publicly available quantitative analysis on the impact of reusable Actions on security vulnerabilities in software projects.
This shows that there is an urgent need for further research as well as appropriate tooling to support developers of reusable Actions and workflows in assessing and hardening their security. A first step in this direction is \github's built-in dependency monitoring service \texttt{Dependabot} which has started to support \gha workflows in January 2022 and reusable Actions in August 2022.\footnote{\url{https://github.blog/2022-08-09-dependabot-now-alerts-for-vulnerable-github-actions/}}

\medskip \noindent \textbf{Abandonment and deprecation.}
Another important challenge that packaging ecosystems face is the risk of packages becoming unmaintained or deprecated~\cite{cogo2021deprecation} when some or all of their core contributors have abandoned the package development~\cite{Constantinou2017,Avelino2019,Kaur2022}. If this happens, the packages may become inactive, implying that bugs and security vulnerabilities will no longer be fixed. This will propagate to dependent packages that rely on such packages.
Cogo \etal~\cite{cogo2021deprecation} have studied the phenomenon of package deprecation in the npm packaging ecosystem, observing that
$3.2\%$ of all releases are deprecated, $3.7\%$ of the packages have at least one deprecated release, and $66\%$ of the packages with deprecated releases are fully deprecated.
Constantinou \etal~\cite{Constantinou2017} studied the phenomena of developer abandonment in the RubyGems and npm packaging ecosystems to determine the characteristics that lead to a higher probability of abandoning the ecosystem.
Developers were found to present such a higher risk if they: do not engage in discussions with other developers, do not have strong social and technical activity intensity, communicate or commit less frequently, and do not participate in both technical and social activities for long periods of time.
Avelino \etal~\cite{Avelino2019} carried out a mixed-methods study to investigate project abandonment in popular \github projects, revealing that some projects recovered from the abandonment of key developers because they were taken over by new core maintainers that were aware of the project abandonment risks and had a clear incentive for the project to survive.

Since Actions are reusable software components being developed in \github repositories, the \gha ecosystem is likely to suffer from this risk of abandoning developers and the presence of unmaintained or obsolete Actions. This calls for studies to quantify this phenomenon and mechanisms to avoid abandonment or to provide solutions to overcome the negative effects of such abandonment. Examples of such solutions could be: finding the right replacement for abandoning developers in Action repositories or suggesting consumers of unmaintained Actions to migrate to alternative Actions.

\begin{figure}[!h]
  \centering
  \includegraphics[scale=0.7]{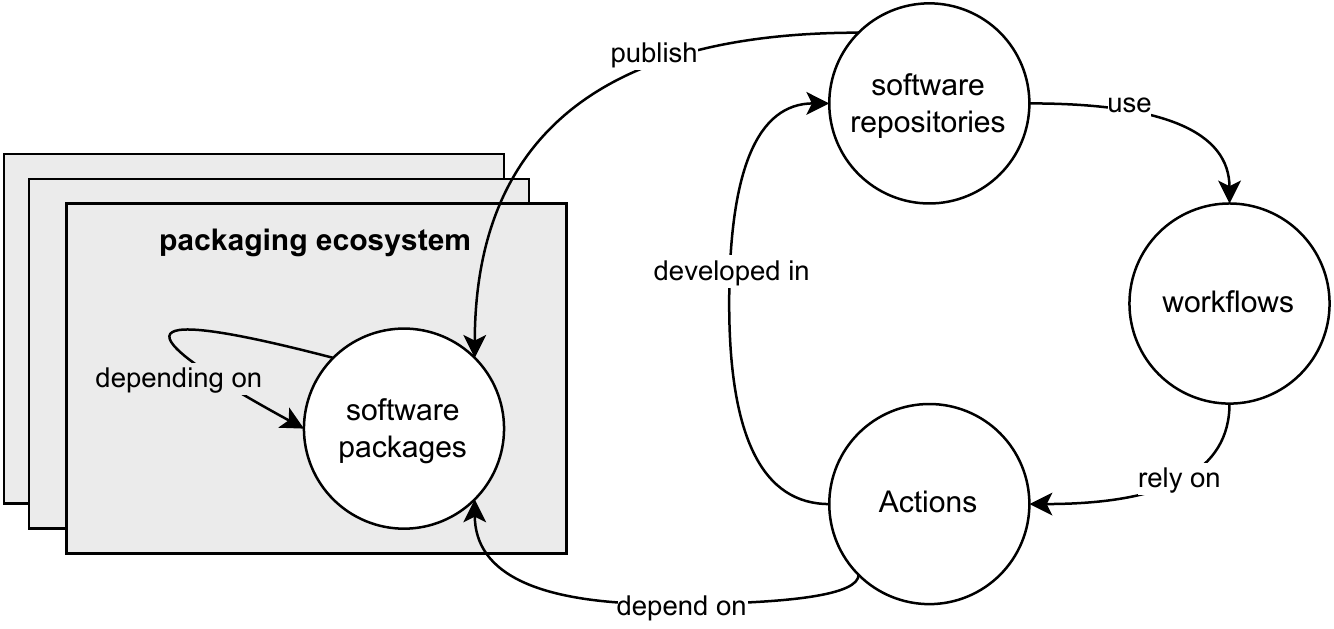}
  \caption{Interweaving of the \gha ecosystem and software packaging ecosystems.}
  \label{fig:actions_ecosystem}
\end{figure}

\medskip \noindent \textbf{Beyond \gha.}
The exposure of \gha to the well-known issues that packaging ecosystems face is all the more worrying because they are not limited to the \gha ecosystem but may also affect other packaging ecosystems. Conversely, the \gha ecosystem may be affected by issues coming from packaging ecosystems.
This situation is depicted in Figure~\ref{fig:actions_ecosystem}:
\github hosts the development repositories of many software projects distributed in packaging ecosystems.
These development repositories may define automated workflows relying on reusable Actions.
The Actions themselves are also developed in (and directly accessed through) \github repositories.
Since Actions are software components developed in some programming language (mostly in TypeScript currently), they may depend on reusable packages or libraries distributed in package registries such as npm.

This potentially strong interconnection between \gha and packaging ecosystems is not without practical consequences given the issues that these ecosystems may face.
Instead of being mostly limited to their \emph{own} ecosystem, issues affecting either packages or Actions may cross the boundaries and propagate to the other software ecosystems they are interwoven with.

Consider for example a reusable Action affected by a security vulnerability. To start with, this vulnerability may compromise all the workflows relying on the affected Action. Next, it may also compromise the development repositories in which these workflows are executed. By extension, it may also affect all the software projects developed in these repositories. In turn, these projects may affect all the dependent packages that use them, and so on.
For example, the \texttt{action-download-artifact} Action, used by several thousands of repositories, was found to expose workflows using it to code injection attacks.\footnote{\url{https://www.legitsecurity.com/blog/github-actions-that-open-the-door-to-cicd-pipeline-attacks}}
Conversely, Actions may depend on vulnerable packages distributed in a packaging ecosystem such as npm. As a consequence, issues affecting these packages may propagate to the Actions using them, and may in turn propagate to the workflows and development repositories relying on these Actions.

\medskip
In summary, many of the issues that software packaging ecosystems have been shown to face also apply directly or indirectly to the \gha ecosystem. Even worse, given that both kinds of ecosystems are tightly interwoven, issues in either ecosystem can and will propagate across ecosystem boundaries, which may lead to a significantly increased exposure to vulnerabilities and other socio-technical health issues.
This raises the urgent need to conduct empirical research for understanding
the extent of these issues, analyse their impact and propagation, and provide tool support for helping repository, package and workflow maintainers.

\section{Discussion}
\label{WFA:sec:discussion}

This chapter focused on the emerging ecosystems of development workflow automation in the \github social coding platform, consisting of the socio-technical interaction with  bots  (automated software development agents), and the workflow automation offered through \actions.
Combined together, \github's socio-technical ecosystem comprises human contributors, bots, workflows and reusable actions, \github Apps\footnote{\url{https://docs.github.com/en/developers/apps}}
 and all of the \github repositories in which these technologies are being developed and used. It also comprises external CI/CD services or other development automation tools that may be used by these \github repositories.
In addition to this, there is a tight interweaving with software packaging ecosystems, since software packages may be developed using bots and \gha, and the development of bots and Actions may depend on software packages.
We have argued that the intricate combination of workflow automation solutions constitutes an important and increasing risk that exposes the involved  repositories -- and by extension the software products they generate or that depend on them -- to vulnerabilities and other socio-technical issues.
Similar issues are likely to apply to other social coding platforms (\eg GitLab and BitBucket) for the same reasons as in \github, even though the workflow automation solutions and technologies in those platforms may be different.

We also argued that bots play an important role in the \emph{social} fabric of the \github ecosystem, since bots interact and communicate with human contributors using a similar interface as the one used by humans (\eg posting and reacting to comments on issues, PRs, code reviews, and commits in repositories).
Actions, on the other hand, are more commonly used to automate \emph{technical} tasks such as executing test suites, deploying packages. This was quantitatively observed by Decan \etal in~\cite{decanuse}.

However, the boundaries between the bot ecosystem and the \actions ecosystem are becoming more and more diffuse.
For instance, nothing prevents bots from directly using the functionality offered by Actions (e.g., a bot could trigger the execution of a workflow using Actions that runs test suites).
Similarly, an Action may instruct a bot to interact with developers and users (e.g., a code coverage Action may report its results through some \github badge, issue or PR comment).

Existing workflow automation solutions were already offered through a wide variety of channels for \github, for example through CI/CD services, external bots, dedicated web interfaces, or \github Apps. The introduction of \actions has further increased the overlap between the possible automation services.
For instance, some automation services that used to be offered through bots or \github Apps have now become available as Actions as well.
An example is the \github App \texttt{the-welcome-bot} for welcoming newcomers, a task for which more recently a \github Action \texttt{wow-actions/welcome} has become available.
Two other examples are the \texttt{renovate} dependency update service and the \texttt{codecov} code coverage analysis that used to be available through web services and \github Apps, and the latter has become offered as a \github Action as well.
Going one step further, \texttt{dependabot}, which used to be an independent bot service, has now become fully integrated into the \github platform.

All these examples illustrate that bots, Apps, Actions and external services will continue to co-exist side by side as part of the development workflow ecosystem.
It is yet unclear to which extent \github repositories are using a combination of workflow automation solutions, or to which extent they tend to migrate from one solution to another.
Hence, empirical studies that shed a deeper insight into this rapidly expanding ecosystem are urgently needed.

\bigskip\noindent\textbf{Acknowledgments.}
This work is supported by the ARC-21/25 UMONS3 Action de Recherche Concertée financée par le Ministère de la Communauté française - Direction générale de l’Enseignement non obligatoire et de la Recherche scientifique, as well as by the Fonds de la Recherche Scientifique - FNRS under grant numbers O.0157.18F-RG43, T.0149.22 and F.4515.23.

\bibliographystyle{spmpsci}
\bibliography{book/references.bib}

\end{document}